# Use of fitted polynomials for the decentralized estimation of network variables in unbalanced radial LV feeders

Valentin Rigoni, Alireza Soroudi and Andrew Keane

School of Electrical and Electronic Engineering, University College Dublin, Belfield, Dublin, Ireland
valentin.rigoni@ucdconnect.ie

**Abstract:** The lack of comprehensive monitoring equipment in low voltage (LV) residential feeders, impedes a near-term deployment of centralized schemes for the integration of domestic-scale distributed generation (DG). In this context, this paper introduces a technique that generates a set of fitted polynomials, derived from offline simulations and regression analysis, that characterise the magnitude of representative network variables (i.e. key for network operation) as a direct analytical expression of the controllable local conditions of any DG unit (i.e. active and reactive power injections). Crucially, the coefficients of these polynomials can be estimated, autonomously at the location of each DG unit, without the need for remote monitoring (i.e. using only locally available measurements). During online implementation, the method only consists of direct calculations (i.e. non-iterative), facilitating real-time operation. The accuracy of the polynomials to estimate the magnitude of the network variables is assessed under multiple scenarios on a representative radial LV feeder. Furthermore, the robustness of the method is demonstrated under the presence of new generation and electric vehicles.

## 1. Nomenclature

| | |
|---|---|
| $a_{X,ij}$ | polynomial coefficient for $b_{X,i} = f(V_l^*)$ |
| $b_{X,i}$ | polynomial coefficient for $X = f(P_{level}, PF)$ |
| $\widehat{b_{X,i_k}}$ | predicted value of $b_{X,i}$ for the $k^{th}$ demand scenario |
| $\mathbf{C}_k$ | $k^{th}$ cluster |
| $\mathbf{c}_k$ | centroid of cluster $\mathbf{C}_k$ |
| $H$ | set of network customers |
| $h$ | network customer |
| $I_{flow}$ | current flow magnitude at any branch |
| $\mathbf{l}_{s,t}$ | normalised pattern for each demand scenario |
| $K$ | number of representative demand scenarios |
| $l_{s,h,t}$ | normalised values of $v_{s,h,t}$ |
| $\mathbf{L}_t$ | set of normalized demand scenarios patterns |
| $P_{level}$ | system active power generation level |
| $PF$ | DG units power factor |
| $\mathbf{p}_{s,t}$ | pattern characterizing each demand scenario |
| $\mathbf{P}_t$ | set of demand scenarios patterns |
| $R^2$ | correlation coefficient |
| $S$ | set of Monte Carlo demand scenarios |
| $s$ | demand scenario in offline Monte Carlo simulations |
| $\sigma_S$ | standard deviation of the distribution of the mean of $S$ samples |
| $\sigma_{all}$ | standard deviation of the distribution of the mean of $S$ samples with $S \to \infty$ |
| $SPC$ | within-cluster sums of pattern-to-centroid distances |
| $t$ | time step in offline Monte Carlo simulations |
| $V_l$ | local CPOC voltage |
| $V_l^*$ | reference local CPOC voltage |
| $v_{s,h,t}$ | CPOC voltages for Monte Carlo demand scenario |
| $X$ | network variable (e.g. nodal voltage magnitude) |
| $x_n$ | calculated value of dependent variable |
| $\hat{x}_n$ | predicted value of dependent variable |

## 2. Introduction

The adoption of domestic-scale distributed generation (DG) can lead to the violation of voltage and thermal limits when the hosting capacity of low voltage (LV) distribution feeders is exceeded [1]. Consequently, it is likely for distribution network operators (DNOs) to implement strategies for the active regulation of DG's active and reactive power outputs [2], raising the need for scalable cost-effective active network management solutions that can be implemented in the near-term.

Active regulation of DG operational setpoints is a mechanism that has been widely explored in the literature [3]. While the categorization of the proposed approaches can be based on various aspects (e.g. need of network topology, consideration of uncertainties, etc.), the required level of network coverage is crucial when assessing their practicality [4]. Within this aspect, methods can be divided into centralized and decentralized solutions [5]. In the case of centralized solutions, all decisions and processes are handled at a top-level [6], [7]. The problem is that such approach heavily relies on the existence of comprehensive and advanced telemetry, which is not present in most distribution systems [4], [5]. Therefore, in recent years, decentralized solutions have been proposed as a cost-effective alternative [8]–[10]. Differently from centralized schemes, they operate control units autonomously and take distributed actions across the network without complete telemetry.

Regardless of the solution approach (i.e. centralized vs decentralized), a concept underlying the active regulation of DG is the need for network observability (i.e. knowledge of the real-time magnitude of the network variables), which is key for identifying the network operational conditions [11]. At the transmission system, observability is achieved by means of State Estimation (SE) [11]. SE is used by operators to infer the state of the system based on measurements scattered across the network. With the advent of DG and the inherent need for active network management strategies, SE has been proposed to provide DNOs with network observability. However, despite classical SE being consolidated at the transmission level, its implementation at the distribution level is unlikely to happen [12]. For instance, classical SE requires access to a degree of monitoring and communication that is usually inexistent at the distribution grid. While it is expected for monitoring capabilities to increase at distribution feeders with the implementation of smart grid technologies [13], global deployment of advanced telemetry will not be immediate and may be avoided due to cost. Furthermore, even if having access to widespread measurements, SE may well prove incompatible with real-time operation [12]. These challenges have driven the development of Distribution State Estimation (DSE) algorithms, tailored for distribution systems.

**Table 1** Overview of the pros and cons of different methodologies to achieve network observability

| Method | Reference | Type of approach | Pros | Cons |
|---|---|---|---|---|
| State Estimation | [11] | Centralized | • Including general measurement functions<br>• Identifying bad data through monitoring redundancy | • Requires widespread monitoring and communication |
| Distribution State Estimation | [14]-[20] | Centralized | • Same benefits as State Estimation<br>• Enhanced convergence and consistency compared with State Estimation | • Requires widespread monitoring and communication (less than State Estimation) |
| Estimation using head of the feeder measurements | [21]-[22] | Decentralized | • No need for remote monitoring | • Active management actions are limited to the transformer location |
| Fitted polynomials | [4],[23],[24] | Decentralized | • No need for remote monitoring<br>• DG units are provided with direct analytical expressions relating network variables with DG active/reactive power injections | • Impossible to identify bad data (no monitoring redundancy)<br>• Polynomials are associated with a specific network configuration<br>• Limited to balanced systems with a few DG units or for estimating sensitivity coefficients |
| Fitted polynomials | Proposed methodology | Decentralized | • Same benefits as [4], [23] and [24]<br>• Can account for multiple DG units<br>• Suitable for unbalanced LV systems<br>• Considers non-uniform demand variations | • Impossible to identify bad data (no monitoring redundancy)<br>• Polynomials are associated with a specific network configuration |

First instalments of DSE can be referred to [14], [15], which aim to enhance the computational efficiency of conventional least-squares estimation (LSE), account for three-phase unbalanced formulations and explore minimum real-time data requirements. More recent DSE formulations can be found in [16]–[20]. With a centralized approach, effort has been made on enhancing LSE's convergence and consistency and on dealing with partial lack of measurements. For instance, the authors in [19] introduce an improved admittance matrix-based state estimator for medium voltage (MV) that simplifies the modelling of zero injections and improves convergence and statistical consistency. In [16], load monitoring in LV feeders is substituted by pseudo-measurements derived from statistical demand models to achieve the condition of observability. The model in [17] uses Gaussian mixtures to represent measurements with non-normal probability densities, claiming that a specific distribution cannot be always considered. A linearized formulation for the DSE algorithm is introduced in [18] which presents lower computational burden and improves convergence. Finally, the work in [20] tackles the lack of synchronism that a widespread LV monitoring infrastructure, if set up, would encounter.

While recent DSE methodologies should facilitate real-time observability in distribution systems, they still rely on widespread telemetry, requiring measurement synchronization and communication between DG units and a central coordinator (e.g. SCADA system). As such, there is no assurance to the DNO that estimations and consequent DG setpoints will perform as expected if deploying these techniques without the assumed telemetry in place. Therefore, in a less conventional way, recent publications have attempted to achieve observability without a communication medium. The authors in [21] and [22] propose a way of estimating remote voltages in LV feeders by means of power flow measurements at the distribution transformer location. Estimations have shown to be highly accurate but autonomous decisions are limited to potential voltage regulation from on-load-tap-changer fitted transformers. Differently, fitted polynomials have enabled to estimate the magnitude, [23], or sensitivities, [4], [24], of network variables using local measurements. Nonetheless, as the methods in [23] and [4] were designed having in mind balanced MV distribution systems with a few wind farms, they cannot be directly implemented at the LV level due to:

- the number of DG units in an LV network can be much larger, requiring different assumptions in terms of active and reactive power injections.
- LV feeders are highly unbalanced.
- residential loads do not vary uniformly.

Finally, the sensitivity analysis in [24] accounts for some of these aspects but does not address the estimation of the network variables' magnitude. An overview of all the mentioned previous publications is shown in Table 1.

Here, the following question arises: can domestic-scale DG units be provided with direct analytical expressions that characterise the magnitude of network variables as a function of their operational setpoints? In response, this paper presents a methodology that uses offline simulations and regression analysis to obtain a set of fitted polynomials that express the magnitude of the representative network variables (i.e. key for network operation) of a radial LV feeder as a function of its DG units operational setpoints. Crucially, the coefficients of these polynomials can be estimated at the location of any DG unit using locally available measurements.

The main contributions of this paper can be summarized as:

- Obtaining direct analytical expressions that relate the magnitude of multiple representative network variables with the operational setpoints of the DG units in a radial unbalanced LV feeder
- Autonomous (i.e. fully decentralized) estimations of the magnitude and sensitivities of the network variables from the location of any DG unit.

Content is structured as follows: Section 3 introduces the methodology. Section 4 describes the real unbalanced LV feeder with photovoltaics (PV) used as a study case. Section 5 includes a multi-scenario validation of the proposed polynomials. Finally, conclusions take place in Section 6.

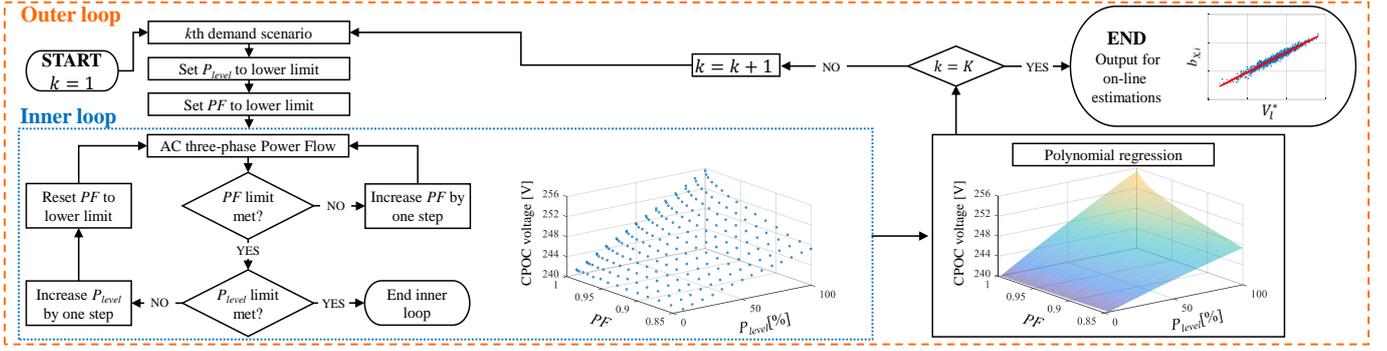

***Fig. 1.*** *Offline stage of the methodology: power flow calculations for system characterization and pre-computed polynomials*

## 3. Methodology

In the proposed method, polynomial equations are used to fit the calculated voltages, flows and losses from a series of offline power flow simulations as a function of the measurements accessible at each DG unit location. These polynomials enable to estimate, from the location of each DG unit, the magnitude of remote network variables and their sensitivities to the DG units setpoint.

### 3.1. Offline stage: power flow calculations and fitted polynomials

The offline stage of the method comprises a series of power flow simulations and polynomial regression. It takes place prior to online implementation and consists of characterizing the evolution of relevant network variables, under different demand scenarios and multiple DG units operational setpoints, with polynomial functions. The overall procedure is outlined in Fig. 1.

*3.1.1. Power flow calculations and data gathering:* The data gathering process involves a series of power flow calculations that create a data set for the training of a regression model. These calculations take place during the inner loop in Fig. 1. Before the beginning of the loop, a demand scenario, that corresponds to a predefined allocation of the demand of each load, is defined and remains fixed. Then, a series of AC unbalanced power flows are solved at different combinations of the DG units operational setpoints. These setpoints are defined by two variables: the system active power generation level, $P_{level}$, and the DG units power factor, $PF$; with $P_{level}$ defined as the ratio, in percentage, between the active power output and the kW rating of the DG units. All $PF|P_{level}$ combinations are obtained by varying them in regular steps between their bounds. This is exemplified in Fig. 1 by a graph showing the voltages (dots) calculated at a generic CPOC.

Despite maintaining the same $PF|P_{level}$ steps, the power flow calculations from the inner loop will change according to the power consumption of the customers in the system. Therefore, as part of the offline stage, an outer loop iterates over a set of $K$ representative demand scenarios with a predefined allocation of residential load demand; this enables the evolution of the inner loop calculations, under multiple diverse demand allocations, to be captured.

*3.1.2. Representative demand scenarios*: A total of $S$ Monte Carlo (MC) simulations are used to explore the system under multiple demand scenarios. Each simulation is obtained by randomly allocating a demand profile, from a statistical model or historical data, to every load in the feeder. Then, each $s$ scenario (for every time step, $t$, considered during the offline methodology) is characterized by a pattern, $\mathbf{p}_{s,t}$, composed of the calculated CPOC voltages, $v_{s,h,t}$, of all $H$ customers. This results in the data set $\boldsymbol{P}_t$:

$$\boldsymbol{P}_t = \{\mathbf{p}_{s,t} = \{v_{s,h,t}\}, s = 1, \dots, S; h = 1, \dots, H\} \quad (1)$$

A normalization process is then used to transform the original sets into the normalized data sets, $\boldsymbol{L}_t$:

$$\boldsymbol{L}_t = \{\mathbf{l}_{s,t} = \{l_{s,h,t}\}, s = 1, \dots, S; h = 1, \dots, H\} \quad (2)$$

$$l_{s,h,t} = \frac{v_{s,h,t} - min_{s=1,\dots,S}\{v_{s,h,t}\}}{max_{s=1,\dots,S}\{v_{s,h,t}\} - min_{s=1,\dots,S}\{v_{s,h,t}\}} \quad (3)$$

Which entries are obtained from the min-max normalization formula (3), that scales the entries of each pattern between zero and one [25]; where $min_{s=1,\dots,S}\{v_{s,h,t}\}$ and $max_{s=1,\dots,S}\{v_{s,h,t}\}$ denote minimum and maximum values.

The obtained normalized sets feed a clustering methodology that allows defining the representative demand scenarios required for the outer loop iterations [26]. Clustering algorithms enable grouping all patterns (i.e. MC scenarios) within each set $\boldsymbol{L}_t$ into clusters (i.e. groups) according to their similarity.

Two clustering algorithms are considered, namely: hierarchical clustering and k-means++ [26]. The first is an agglomerative deterministic method for which both Ward's and average methods are accounted for. The second is heuristic and more effective than traditional k-means. The result is a set of $K$ clusters of similar patterns, each one defined by a centroid equal to the average of all elements belonging to the cluster. Every centroid leads to a representative $k^{th}$ demand scenario associated with the pattern, with the smallest Euclidean distance to it. The representative scenario probability $\omega_k$ equals the ratio between the size of cluster $k$ and that of the original data set.

*3.1.3. Regression analysis - polynomial fitting:* At the end of every inner loop, the compilation of calculated values for each variable $X$ under interest (e.g. CPOC voltages, flows, losses, etc.) is characterized by the second-degree polynomial from Equation (4) (see Fig. 1), where $\tau = \tan(\text{acos}(PF))$:

$$\begin{aligned} X = {} & b_{X,1} + b_{X,2}\, P_{level} + b_{X,3}\, \tau + b_{X,4}\, P_{level}^{2} \\ & + b_{X,5}\, P_{level}\, \tau + b_{X,6}\, \tau^{2} \quad (4) \end{aligned}$$

$$\min \sum_{n=1}^{N} (\hat{x}_n - x_n)^2 \quad (5)$$

Equation (4) characterizes every variable $X$ with a direct analytical equation of $P_{level}$ and $PF$, i.e. $X = f(P_{level}, PF)$, with the latter representing the active and reactive power injected by the DG units in the network.

For each dependent variable $X$, the determination of the six coefficients ($b_{X,i}$) is done using polynomial regression [27]. This consists of minimizing the Sum of Squared Residuals (SSR) as per equation (5); with residuals defined as the difference between the values of the dependent variable predicted by the model (4), $\hat{x}_n$, and the ones calculated during the inner loop, $x_n$. In (5), $N$ is the total number of samples (i.e. $P_{level}|PF$ permutations).

The demand allocation for each of the executions of the inner loop in Fig. 1 will affect the evolution of the variables under interest (as the results of the power flow calculations will depend on the system demand). Hence, each one of the $K$ representative scenarios results in different values for the coefficients for (4). This means that during online implementation, at each DG unit location, the polynomial (4) that better characterizes the network under the current demand (unknown by the DG unit) needs to be identified. Therefore, in the offline stage, a second polynomial regression expresses all the coefficients in (4) from all demand scenarios as a linear equation of a reference local voltage $V_l^*$ for each DG unit (see Fig. 1; end box):

$$b_{X,i} = a_{X,i1} V_l^* + a_{X,i2} \quad (6)$$

where $V_l^*$ is the single-phase voltage at the unit's CPOC; calculated in the inner loop with $PF$ and $P_{level}$ at their lower bounds. Equation (6) is crucial for the method's real-time implementation. As shown later, it allows to estimate (at each DG unit location) every $b_{X,i}$ in (4) based on the local voltage. The coefficients $a_{X,ij}$ in (6), for each $b_{X,i}$ of every variable $X$, are obtained by solving a weighted SSR minimization:

$$\min \sum_{k=1}^{K} \omega_k \left(\widehat{b_{X,i_k}} - b_{X,i_k}\right)^2 \quad (7)$$

where $K$ is the total number of scenarios, $\omega_k$ the probability of the $k^{th}$ scenario, $b_{X,i_k}$ the previously calculated values for the coefficients in (4) and $\widehat{b_{X,i_k}}$ the ones estimated by (6).

The described offline simulations are used to characterize the feasible space of all relevant variables $X$ under different DG units operational setpoints and demand scenarios. While previous decentralized methodologies, [28], [29], neglect demand seasonal behaviour [30]. Here, the coefficients in (4) and (6) depend on time (as the representative demand scenarios are time-dependent). Note that, for simplicity, the polynomial coefficients are not shown with index $t$.

### 3.2. Online stage: Estimating the most likely polynomials based on real-time local measurements

Once the offline calculations in Fig 1 are finished, each DG unit is provided with a set of equations in the form

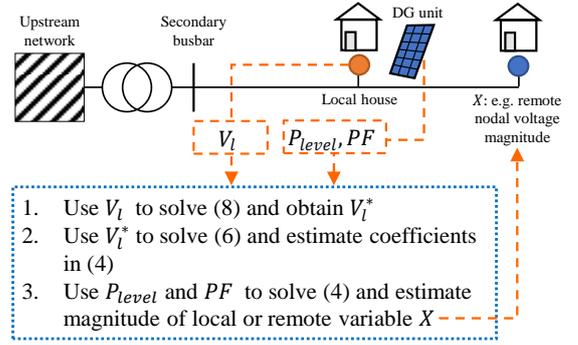

Fig. 2. Summary of the calculations for online implementation

$$V_l^* = \{V_l - [a_{V_l,12} + a_{V_l,22} P_{level} + a_{V_l,32} \tau + a_{V_l,42} P_{level}^2 \\ + a_{V_l,52} P_{level} \tau + a_{V_l,6} \tau^2]\}/\{a_{V_l,11} \\ + a_{V_l,21} P_{level} + a_{V_l,31} \tau + a_{V_l,41} P_{level}^2 \\ + a_{V_l,51} P_{level} \tau + a_{V_l,61} \tau^2\} \quad (8)$$

of (6), dependent on its location, for the variables under interest. During online implementation, estimations at each DG unit will be obtained autonomously following the process described below. Note that, differently from a centralized approach, DG units do not communicate with each other or with a central coordinator.

The online process, summarized in Fig. 2, only requires the local measurement of the CPOC voltage and the DG unit active and reactive power setpoints. It consists of a limited series of direct calculations that take place autonomously at each DG unit location. First, the local voltage, $V_l$, is transformed into the reference voltage $V_l^*$ from (6). During the offline stage, the latter corresponded to the voltage at the local CPOC calculated with both $P_{level}$ and $PF$ at their bounds. Given that these two parameters will not necessarily be at their bounds during online operation, a simple direct calculation is needed. as in equation (8). The expression in (8) can be obtained when solving the polynomial (4) associated with the local voltage $V_l$, i.e. variable $X$, for $V_l^*$. (with all $b_{V_l,1}$ replaced by the corresponding model in (6)). Then, with the value of $V_l^*$, the appropriate coefficients in (6) are used to estimate the values for every $b_{X,i}$ in (4).

Once the previous calculations are performed, the set of polynomials (4) that better correspond to the current system demand is available at the specific DG location. On the one hand, these polynomials can be used to estimate the magnitude of remote voltages, losses and branch flows by simply entering the current value for $PF$ and $P_{level}$ in (4) (like exemplified in Fig. 2). This allows to autonomously achieve network observability at each DG unit location. On the other hand, if $PF$ and $P_{level}$ are regulated, e.g. due to a voltage violation, the polynomials can be used to verify that the new expected voltages are within the desired limits. This can be done as the expression in (4) explicitly contains the variables' sensitivities to the DG setpoints. As the described process consists of simple non-iterative calculations, it poses very low computational burden with no risk of non-convergence during real-time operation. Furthermore, as it is carried out autonomously, no centralized data-acquisition and decision-making scheme is needed (e.g. SCADA system); enabling each DG unit to self-monitor its actions without a communication medium.

## 4. Test system

The methodology is implemented on a real LV feeder, selected from a set of networks from [31]. This selection is based on this feeder showing similar characteristics to those of the most typical feeders that are expected to present technical problems due to DG [32]. It has a nominal line voltage of 400 V and consists of 1,684 metres of 4-wire cable and 83 single-phase customers and is illustrated in Fig. 3.

For this test case (unless otherwise stated) a 100% PV penetration, i.e. every house has a PV, is considered (a worst-case concerning technical impacts). All PVs are modelled as single-phase constant PQ generators, and their ratings are allocated based on historical data [33], with values that vary from 1.0 to 4.0 kW. In addition, the load model from [30] is used to obtain the representative scenarios for the demand allocation. Finally, the system active power generation level, $P_{level}$, and the power factor, $PF$, are assumed to vary uniformly, during the offline computations, for all DG units. A uniform $P_{level}$ means that all units operate at the same percentage of their rated active power. This assumption is very common within decentralised techniques, and is funded on the limited geographical extension of LV residential feeders [21], [28], [29].

For the inner loop offline calculations in Fig. 1, the upper and lower limits for $P_{level}$ and $PF$ are defined based on the expected real-time operation of the PV systems. As $P_{level}$ is defined as the ratio, in percentage, between the active power output and the kW rating of the PV units, its value is bounded by zero and 100%. The power factor, $PF$, is bounded between 1 and 0.85 (inductive). The lower limit for $PF$ should not be too low (the Irish distribution code defines a minimum power factor of 0.95 [34]). An unjustified low power factor limit will result in more steps for an adequate fitting and in unnecessary computations. Note that these limits can be different if required by the distribution network code. In terms of the simulation platform, AC power flow simulations are performed in OpenDSS [35]. In addition, regression analysis is performed with MATLAB's least square regression analysis function [36].

## 5. Results & discussion

### 5.1. Offline stage analysis

*5.1.1. Defining the $PF|P_{level}$ step size:* Regression analysis theory states that, for a good predictive performance, the data sample needs to properly describe the trend of the related variables [27]. Therefore, in the inner loop from Fig. 1, the number of observations, $N$, must be adequately defined. These observations are proportional to the number of steps considered when varying $P_{level}$ and $PF$ along their bounds. While increasing the number of steps results in a better characterisation of the variables, it also increments the required power flow simulations, which can extend to unnecessarily high computational times. Here, based on the analysis of the data, $n_{steps} = 20$ was found to be adequate.

*5.1.2. Validation of the polynomial regression models:* The order of the models selected in polynomial regression is empirical and derives from the analysis of the training data [27]. It is important to mention that over-fitted models unnecessarily increase the computational difficulty of the problem and can result in ill-conditioned regressions. In an ill-conditioned regression, solving the regression model will magnify any noise in the training data up to a point where the model predictions are dominated by the latter. On the other hand, under-fitted models may not be able to properly capture the trend of the data. In this subsection, it is quantified how well the models (4) and (6) fit the calculations from the offline stage. In order to do that, we propose to use the $R^2$ correlation coefficient [27]. Its value, which varies from zero to one, can be interpreted as the amount of the sample variation explained by the independent variables. Traditionally, values over 0.5 and 0.7 correspond to moderate and strong correlations respectively.

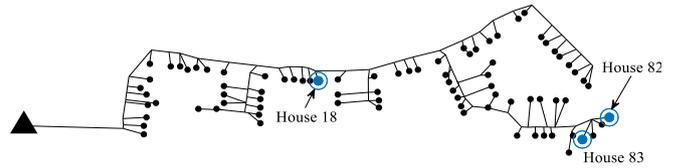

**Fig. 3.** *Real residential representative LV feeder - dots represent houses and the triangle is the head of the feeder*

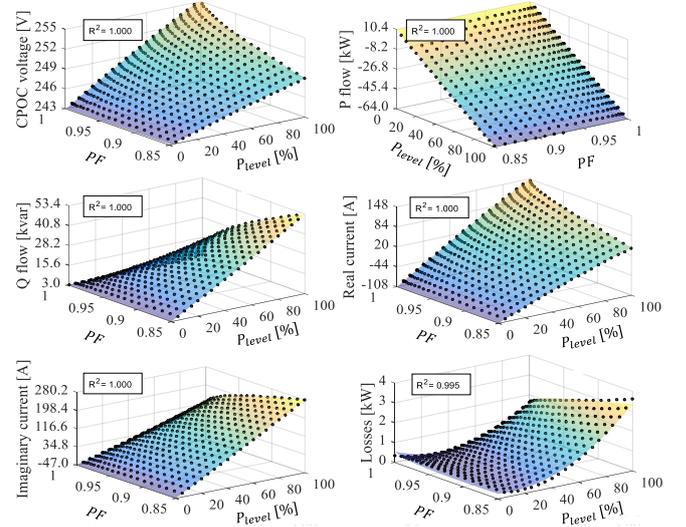

**Fig. 4.** *Model (4) characterization of variables X under interest*

Fig. 4 shows how the polynomial model (4) characterises the variables of interest for a generic demand scenario at 6 pm; where the dots correspond to the values from the power flow simulations. For conciseness, only the CPOC voltage for house 82 (see Fig. 3) is included. This house is of interest as it presents the highest impedance path to the head of the feeder (i.e. it is very exposed to voltage violations). Furthermore, active power losses and flows at the head of the feeder correspond to those in phase *c*. It can be observed that the values obtained for $R^2$ are very close to unity independently of the variable; this means a good fitness of the model. In this method, current flow phasors are decomposed into their real and imaginary parts because a direct treatment of the current magnitudes can result in the inadequate fitting obtained in [23]. Consequently, the magnitude of the current flows, $I_{flow}$, is characterized as:

$$\|I_{flow}\| = \sqrt{[Re(I_{flow})]^2 + [Im(I_{flow})]^2} \quad (9)$$

The minimum value of $R^2$ associated with all representative demand scenarios during a 24-hour timeframe for the full set of variables (i.e. considering all houses and phases) is presented in Table 2. Results for different degrees of the polynomial surface model (4) are also included for comparison purposes. From these results, it is observed that while the goodness of fit improves significantly from a first-degree to a second-degree model, the benefit of increasing the degree to three is negligible. Having a higher degree would increase computational times during the offline stage (as 10 coefficients would need to be obtained instead of 6) and would result in more complex expressions for future applications without providing any clear benefit. Note that during simulations, no ill-conditioned regression was found for any representative demand scenario for the three polynomial degrees in Table 2.

As expressed in Section 3.1, the values of the $b_{X,i}$ coefficients for the polynomials (4) depend on the system's demand. This can be observed in Fig. 5 where the polynomials obtained for the phase $c$ voltage at house 82 are depicted for two different representative demand scenarios. Therefore, in online operation, the coefficients of the polynomials that best represent the actual system demand are estimated using the local reference voltage $V_l^*$ as in (6). The scatter plots in Fig. 6 display the values of the six $b_{X,i}$ coefficients (i.e. circles) for house 82's voltage from all representative demand scenarios against the local reference voltage $V_l^*$ from house 18 phase $c$ (see Fig. 3). Each circle size is proportional to $\omega_k$ and is weighted accordingly in (7) for the definition of the linear model (6) (i.e. dotted line). The value of $R^2$ is always over 0.65 (up to 0.97) except for $b_{V_l,2}$, for which it drops to 0.46. The prediction of this coefficient may improve if extra monitoring is accessible. However, this would mean extending the considered limitations and losing practical value. Nonetheless, as it will be shown, accurate results are obtained in any case; this can be attributed to a limited variability of the coefficient (maximum difference of less than 4% with respect to the proposed linear model).

Maintaining house 18 as the local reference, the mean and minimum values of $R^2$ for model (6) for multiple variables (on phase $c$) are presented in Table 3. Only a linear correlation is considered, as MATLAB's least squares regression analysis function found higher degree models to be ill-conditioned. While for most coefficients, strong correlations are found, there are exceptions; an insightful analysis of network sensitivities can be pursued from their analysis. For example, low correlations can be found for variables whose sensitivities to active or reactive power are expected to be untraceable. For instance, the fifth coefficient (sensitivities to DG reactive power injections) presents low correlations for active power flows and losses. The observed $R^2$ values remain similar, independently of the local house location, if the CPOC is connected to the same phase as that of variable $X$. Otherwise, the values in Table 3 can decrease below 0.5 as the uncertainty in all phases cannot be captured with single-phase measurements. This is not a limitation of the polynomials and can be overcome with local voltage monitoring in all phases. It will be later shown, from a comparison with a benchmark method, that similar results are expected for any method with similar monitoring restrictions.

Overall, results show that the relationship between the studied electrical variables and the DG units' setpoints is characterizable under any demand scenario by a second-degree polynomial function. Furthermore, local voltage measurements can be used to obtain the most likely coefficients of the previous polynomials that correspond to the actual system demand.

**Table 2** Minimum Pearson correlation for polynomial model (4)

| Degree | Houses voltage | Active power flow | Reactive power flow | Real current | Imaginary current | Active power losses |
|---|---|---|---|---|---|---|
| 1 | 0.859 | 0.981 | 0.847 | 0.844 | 0.798 | 0.244 |
| 2 | 1.000 | 1.000 | 1.000 | 1.000 | 1.000 | 0.989 |
| 3 | 1.000 | 1.000 | 1.000 | 1.000 | 1.000 | 0.999 |

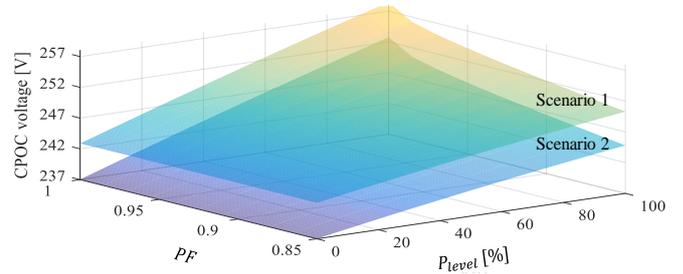

***Fig. 5.*** *Model (4) for remote CPOC voltage at two different demand scenarios*

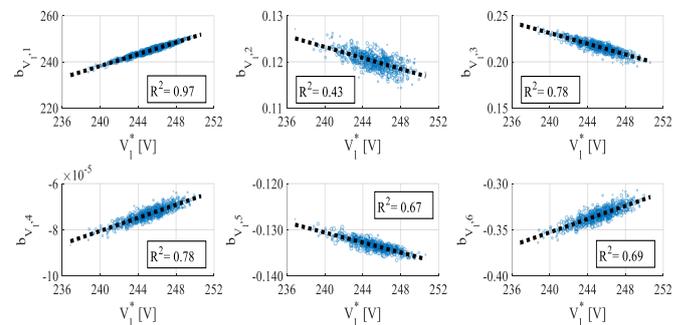

***Fig. 6.*** *House 82 $b_{V_l,i}$ coefficients indicating linear dependency on house 18 phase c reference voltage – 6pm*

**Table 3** Minimum Pearson correlation for polynomial model (6)

| $b_{X,i}$ coefficient | House 82 voltage | | Active power flow | | Reactive power flow | | Real current | | Imaginary current | | Active power losses | |
|---|---|---|---|---|---|---|---|---|---|---|---|---|
| | mean | min | mean | min | mean | min | mean | min | mean | min | mean | min |
| 1 | 0.98 | 0.97 | 0.83 | 0.79 | 0.09 | 0.03 | 0.10 | 0.02 | 0.83 | 0.79 | 0.89 | 0.82 |
| 2 | 0.46 | 0.38 | 0.86 | 0.82 | 0.26 | 0.12 | 0.43 | 0.27 | 0.93 | 0.90 | 0.98 | 0.97 |
| 3 | 0.77 | 0.75 | 0.88 | 0.85 | 0.76 | 0.73 | 0.90 | 0.89 | 0.92 | 0.89 | 0.96 | 0.94 |
| 4 | 0.80 | 0.78 | 0.80 | 0.73 | 0.70 | 0.65 | 0.88 | 0.85 | 0.85 | 0.80 | 0.95 | 0.90 |
| 5 | 0.50 | 0.22 | 0.30 | 0.15 | 0.92 | 0.87 | 0.94 | 0.92 | 0.28 | 0.13 | 0.04 | 0.01 |
| 6 | 0.72 | 0.69 | 0.84 | 0.79 | 0.68 | 0.61 | 0.86 | 0.83 | 0.88 | 0.84 | 0.95 | 0.90 |

*5.1.3. Results from the clustering methodology: representative demand scenarios:* The central limit theorem is used to define a suitable number $S$ of MC simulations for the clustering methodology in Section 3.1. It states that the distribution of the mean of $S$ samples obtained from independent and identically distributed random variables (i.e. demand profiles) is Gaussian if $S \to \infty$ [37], and relates the standard deviation of that distribution, $\sigma_S$, with the one obtained from analysing the whole population, $\sigma_{all}$:

$$\sigma_S = \sigma_{all}/\sqrt{S} \quad (10)$$

Here, 10,000 iterations are considered so that the difference between the results mean and the one for $S \to \infty$ lies within a Gaussian of standard deviation equal to 1% of $\sigma_{all}$.

The best clusters structure is determined by assessing the results obtained from various number of clusters $K$ [38]. For that purpose, the within-cluster sums of pattern-to-centroid distances [26], $SPC$, in (11) was found to be a suitable index; where $\mathbf{c}_k$ is the centroid of cluster $\mathbf{C}_k$. It is a measurement of cluster's compactness and decreases while $K$ increases. Results for the $SPC$ index are depicted in Fig. 7 for all the clustering algorithms. According to this index, both Ward's method and k-means++ outperform the average method. While they show similar results, the first one is preferred as it is deterministic and more time efficient. A knee-point occurs in between $K = 2$ and $K = 100$. The selected number of clusters must be greater than 100 to lead to reliable results. From that point on, the greater the value of $K$, the closer results will be to those from analysing the whole population of 10,000 MC scenarios. As more representative demand scenarios will lead to more calculations, the selection of $K$ must account for the computational times that are acceptable for the DNO. Here, $K = 400$ with Ward's method is considered adequate (i.e. a 96% scenario reduction).

$$SPC = \sum_{k}^{K} \sum_{\mathbf{l}_s \epsilon \mathbf{C}_k} (\|\mathbf{l}_s - \mathbf{c}_k\|_2)^2 \quad (11)$$

*5.2. Online application: estimation of the system variables*

Here, the methodology is applied for the autonomous estimation of local and remote system variables. From this point on, when using the term polynomials, we refer to the model (4); for which the most likely coefficients (that characterize the real-time conditions of the system) have been obtained, autonomously at each PV location, via the measurement of the local CPOC voltage by following the procedure in Section 3.2 (see Fig. 2).

*5.2.1. Assessing the accuracy of the estimations:* A 10-minute resolution time-series daily power flow simulation environment is used to assess the estimations obtained from the polynomials. At each time step, the estimation of every variable $X$ is obtained, autonomously at each PV location, by solving (4). The values of $P_{level}$ derive from a real solar irradiance profile (from an Irish weather station) and customers demand from daily profiles obtained with the statistical demand model from [30]. $PF$ is constant and equal to 0.95 (typical fixed inductive power factor in Ireland [34]). In addition, a decentralized DSE algorithm, solved at each PV location, is used as a benchmark. This DSE serves as a proxy for a traditional LSE formulation; where the network variables are estimated via the iterative solution of a non-convex constraint optimization [11] (more detail on this algorithm can be found in the Appendix). The mean and max differences, in per unit, between the estimation error for both polynomials and DSE are shown in Table 4. It is observed that this difference is very low (less than $1\times10^{-2}$ pu in all cases), meaning that the polynomials truly characterized the power flow equations relating the network variables with the network nodal power injections.

Fig. 8 shows the actual voltage profile, for a generic scenario, for house 82 phase *c* together with the estimations obtained locally from house 18 phase *c*. It can be noted that, due to the presence of PVs, voltages surpass the statutory upper voltage limit of 1.1 pu [34]. The estimations from the proposed methodology result in a good match for the actual voltage profile, with a maximum error of less than $3\times10^{-4}$ pu. On the other hand, Fig. 9 shows the voltage profile for house 83 phase *a* voltage (also located at the end of the feeder) with the estimations obtained from the same previous local house (connected to phase *c*). Now, the phase of the estimated variable is not coincident with the local CPOC. While the trend of the voltage increment due to PV generation is captured, this is not the case of the voltage variations due to changes in demand; this results in a greater gap with respect to the actual profile (of up to 0.02 pu). The similar results from the DSE show that this higher error is not because of the polynomials formulation but due to local voltage monitoring being limited to only one phase, and could be overcome by having access to all three phases in the vicinity to the CPOC if the characteristics of the network allow it.

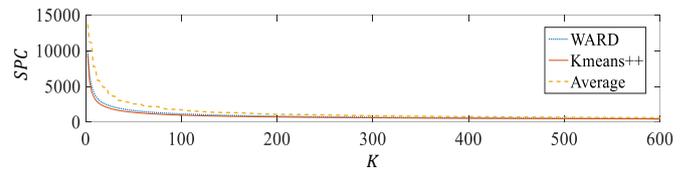

*Fig. 7.* Sum of intra-cluster distance for 2<K<600

**Table 4** Error difference between polynomials and DSE

| Mean difference | House 82 voltage | Active power flow | Reactive power flow | Current | Active power losses |
|---|---|---|---|---|---|
| ΔMean [pu] | $2.0\times10^{-4}$ | $8.4\times10^{-3}$ | $3.6\times10^{-3}$ | $1.6\times10^{-3}$ | $1.0\times10^{-4}$ |
| ΔMax [pu] | $6.0\times10^{-4}$ | $9.1\times10^{-3}$ | $6.5\times10^{-3}$ | $6.4\times10^{-3}$ | $3.0\times10^{-4}$ |

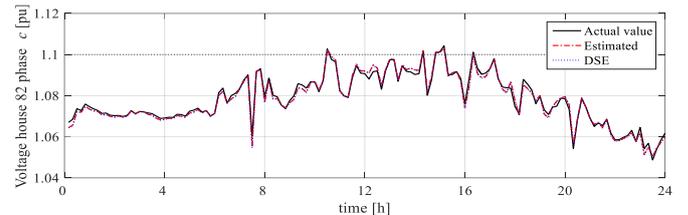

*Fig. 8.* House 82 phase c actual and estimated voltages – Local monitoring at house 18 phase c

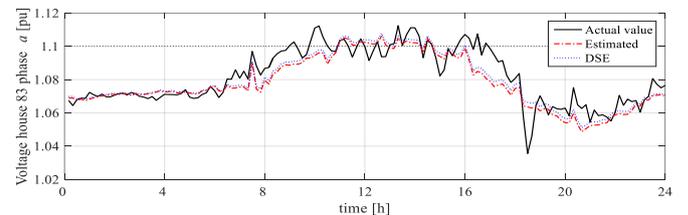

*Fig. 9.* House 83 phase a actual and estimated voltages – Local monitoring at house 18 phase c

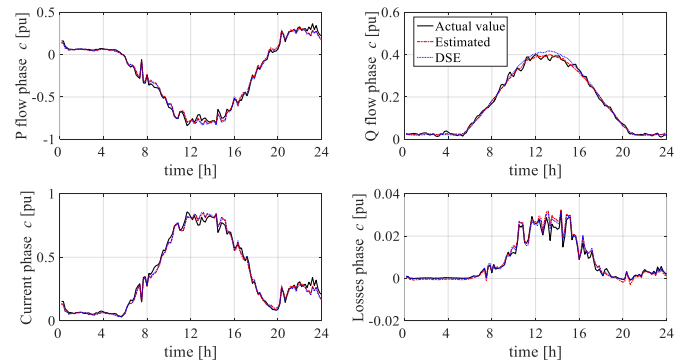

*Fig. 10.* Other phase c estimations – Local monitoring at house 18 phase c

Fig. 10 shows the profiles for phase *c* current/power flows at the head of the feeder and system active power losses. All values are in per unit with a base current of 375 A (main cable rating). Again, the proposed methodology provides good estimations and performs similarly to the DSE. Focusing on reactive power flows, it can be noticed that both polynomials and DSE are unable to capture the random residential load reactive power variations. Nonetheless, the polynomials accurately track the reactive power flows due to PVs inverters reactive power consumption.

To rigorously assess the estimations provided by the polynomials, all time steps for the 400 representative demand scenarios from the clustering procedure are analysed. For instance, Fig. 11 shows actual and estimated values (with measurements at house 18) for house 82 voltage, (a), and head of the feeder phase *c* current, (b). It can be observed that, for both cases, estimated and actual values show an excellent proportionality. A quantification for the quality of the estimates for all variables of interest is presented in Table 5; where the median of the error, the mean and $99.7^{th}$ percentile of the absolute error and $R^2$ are included. The error is quantified as the per-unit difference between actual and estimated values. From these results, it can be observed that:

- The median for all estimates is very close to zero; meaning no systematic under or overestimations.
- Absolute errors present very small values. This applies for the mean and the $99.7^{th}$ percentile of the error distribution.
- Estimated and actual values are strongly correlated ($R^2 \approx 1$).

Finally, to explore how results depend on the location of the local PV, the estimation error for the voltages at the end of the feeder and the currents at the head of the feeder are depicted, considering local measurements at each of the 83 customers, in Fig. 12 and Fig. 13. Results correspond to estimations on coincident phases for each house. While in Fig. 13, the error for current estimations remains almost constant, the error for the voltage estimations in Fig. 12 is affected by the location of the local measurement. Nonetheless, even in the worst case (end of the feeder estimations from head of the feeder customers) it remains small enough to provide reliable estimations.

Overall, the method provides accurate estimations for all the considered relevant variables (i.e. voltages, power/current flows and losses). Combined with its autonomous implementation, this places the methodology as a valid alternative to centralized schemes. In addition, the polynomials showed to be as accurate as an equivalent DSE to estimate the system state; in other words, they truly characterize the power flow equations that relate the system variables with the monitored parameters. Nonetheless, the polynomials have the advantage of providing estimates through non-iterative calculations; this is important as in an online application, DNOs do not need to be concerned about convergency. Furthermore, they also describe, with a direct equation, how the magnitude of the network variables will change if the DG setpoints are modified.

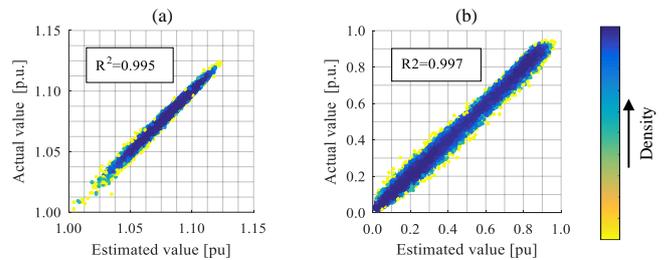

*Fig. 11.* *Actual and estimated house 82 voltage (a) and head of the feeder phase c current flow (b) – Local monitoring at house 18.*

**Table 5** Estimations error and correlation with actual values

| Estimated variable | House 82 voltage | Active power flow | Reactive power flow | Current | Active power losses |
|---|---|---|---|---|---|
| Mean [pu] | $1.06 \times 10^{-3}$ | $2.13 \times 10^{-2}$ | $7.60 \times 10^{-3}$ | $1.72 \times 10^{-2}$ | $1.26 \times 10^{-3}$ |
| Median [pu] | $5.27 \times 10^{-5}$ | $4.85 \times 10^{-3}$ | $-3.76 \times 10^{-4}$ | $9.23 \times 10^{-4}$ | $6.06 \times 10^{-4}$ |
| 99.7 % [pu] | $5.67 \times 10^{-3}$ | $1.01 \times 10^{-1}$ | $3.28 \times 10^{-2}$ | $8.11 \times 10^{-2}$ | $7.33 \times 10^{-3}$ |
| $R^2$ | 0.995 | 0.997 | 0.998 | 0.997 | 0.989 |

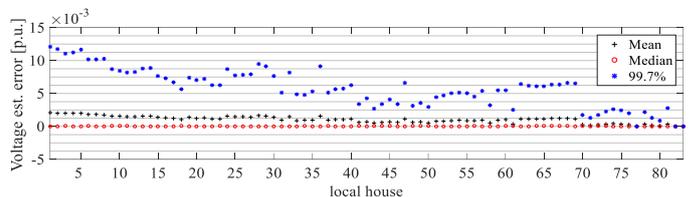

*Fig. 12.* *End of the feeder voltage estimation error – Local monitoring at different houses*

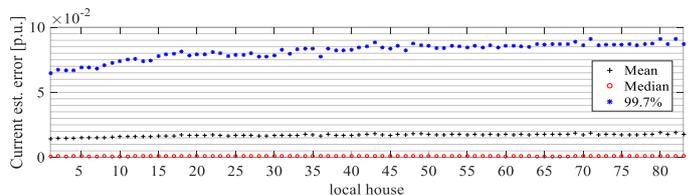

*Fig. 13.* *Head of the feeder current estimation error – Local monitoring at different houses*

**Table 6** Estimations error and correlation with actual values – simulating an error on the local voltage measurement

| Estimated variable | House 82 voltage | Active power flow | Reactive power flow | Current | Active power losses |
|---|---|---|---|---|---|
| Mean [pu] | $7.79 \times 10^{-3}$ | $5.11 \times 10^{-2}$ | $8.17 \times 10^{-3}$ | $3.88 \times 10^{-2}$ | $2.69 \times 10^{-3}$ |
| Median [pu] | $6.52 \times 10^{-5}$ | $6.37 \times 10^{-3}$ | $-3.33 \times 10^{-4}$ | $-2.75 \times 10^{-3}$ | $5.62 \times 10^{-4}$ |
| 99.7 % [pu] | $2.91 \times 10^{-2}$ | $1.93 \times 10^{-1}$ | $3.44 \times 10^{-2}$ | $1.60 \times 10^{-1}$ | $1.76 \times 10^{-2}$ |
| $R^2$ | 0.832 | 0.986 | 0.997 | 0.984 | 0.939 |

*5.2.2. Effect of an error on the local voltage measurement:* Transducer errors will affect the accuracy of the method. To investigate their influence, a Gaussian distributed error, with a $99.7^{th}$ percentile of 2.5%, is added to the local voltage $V_l$ measurement. The 400 representative demand scenarios are explored and results for the estimation error is included in Table 6. As expected, a reduction in the performances observed in Table 5 takes place. For instance, an increment of approximately 0.025 pu in the $99.7^{th}$ percentile for voltage estimations is obtained (as expected from a 2.5% transducer error). However, despite an increment in the observed uncertainties, they remain within an acceptable range.

In contrast to centralized DSE, the proposed method cannot identify bad data (e.g. faulty monitors, wrong phase specification, etc.). The only way of performing full bad data identification is by having redundancy on the measurements scattered across the network [11]. However, such a scenario would require a level of telemetry that exceeds the one that currently exists at the LV level (it is under such context that a decentralised method presents a clear practical value).

*5.2.3. Accuracy of the estimations at different PV penetrations:* All previous results correspond to a 100% PV penetration; in other words, every house has a PV, which is considered as the worst-case from a technical impact perspective. Nonetheless, the performance of the methodology is expected to be robust regardless of the amount of DG. In Fig. 14, the voltages at House 82 phase $c$ are estimated from local monitoring at house 18 over different penetrations of PV, for which the corresponding polynomials have been calculated with the offline methodology in Section 3. It can be observed that, independently of the number of customers with PV, estimations and actual values show to be very similar, with a maximum difference of only 0.0054 pu These results are in line with those obtained in Fig. 8 for a 100% PV penetration.

*5.2.4. Discussion on computational times:* One of the method's benefits is that computational complexity is avoided during online application. Indeed, all the performed estimations took an average time of only 1.6 ms with no risk of non-convergence (i.e. obtained from direct calculations). In terms of the offline simulations, an Intel Xeon E5-2699 allowed obtaining all 10-minute resolution polynomials in 0.79 hrs. This time is low enough for DNOs to feasibly compute the polynomials for a large set of feeders if required. The reason for concentrating computational difficulty in the offline stage, which can be planned in advance, is that the criticality of computational times and convexity takes place during the real-time operation of the system.

*5.2.5. Robustness and adaptability:* compared to a DSE algorithm, a disadvantage of the method is that the coefficients in the polynomials will be associated with a specific network configuration (the one considered during the offline simulations). Consequently, the method's performance is expected to be affected by modifications to network characteristics. While the proposed approach aims to produce an adequate set of polynomials based on the real-time local measurements, the adoption of new DG units, for example, will affect their accuracy. This is due to the absence of those new units during the offline training stage. Nonetheless, this will be unavoidable in any methodology that relies on offline calculations (e.g. volt/var curves).

The adaptability of the obtained polynomials to the presence of new PVs is shown in Fig. 15 and Fig. 16 for voltage and current estimations; for both cases, obtained from house 18. The solid line represents the actual values at a 75% PV penetration. The dashed lines correspond to the estimations obtained from polynomials that were updated at different penetrations; from 50% to 75%. It can be noted that the method is robust with regards to voltage characterizations; where differences between the voltages estimated from the updated and non-updated polynomials are on the order of 0.001 pu. This is important as the DG hosting capacity of LV feeders is mainly limited because of violations to the voltage statutory limits [21]. In the case of the current flows, a systematic underestimation becomes noticeable when there is a significant difference between the original existing PV and the current one. In any case, the proposed approach still leads to promising results and, if not updated, the same polynomials could be used if a more conservative strategy is taken for the branch flows. With respect to the assumption on the uniformity of $P_{level}$ among PVs, this analysis shows that the method is expected to be robust to the presence of clouds and different PV orientations. The latter would also correspond to unexpected different PV injections; probably lower than the ones seen here (with up to a 50% PV capacity mismatch).

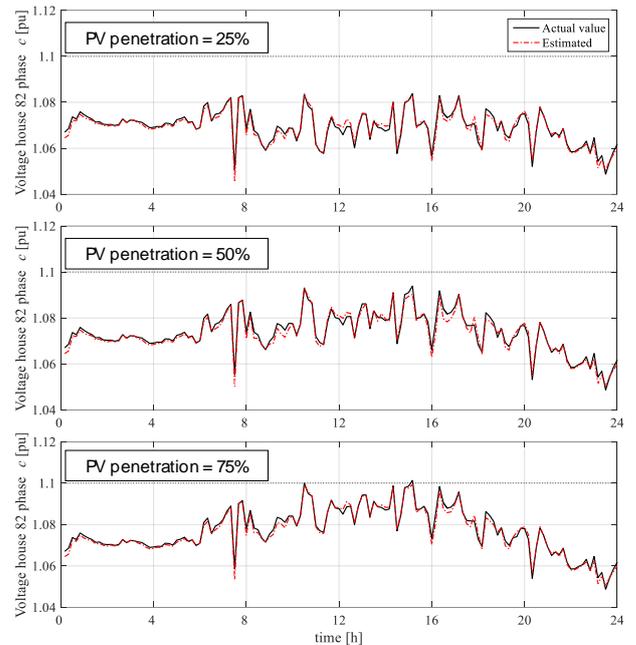

*Fig. 14.* House 82 phase c actual and estimated voltages at multiple PV penetrations (i.e. 25, 50 and 75%) – Local monitoring at house 18 phase c

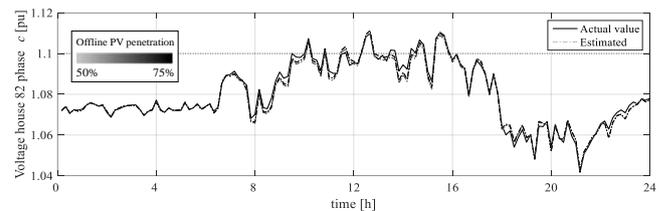

*Fig. 15.* House 82 phase c voltage – Estimations performed with outdated polynomials on a 75% PV penetration scenario

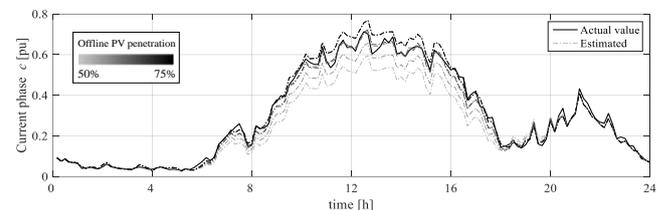

*Fig. 16.* Head of the feeder phase c current – Estimations performed with outdated polynomials on a 75% PV penetration scenario

The demand characteristics can also start differing from the ones considered during the offline stage. For instance, Fig. 17 and Fig. 18 show the same voltage and current estimations in a 50% PV penetration scenario after 1/3 of the customers have adopted electric vehicles (EVs); modelled as 3kW loads using the profiles in [31]. The method is again very robust, showing some current underestimations of less than 0.1 pu at periods of high EV connections (i.e. after 7 pm) but almost no negative effect on the voltage estimations. This also shows that the method is expected to be robust against differences between the load model used for the representative scenarios and reality.

Another aspect to account is the possible change on the network topology (e.g. changes on cable sections and reconfigurations under fault conditions). In such a case, the accuracy of the methodology will also be affected. However, LV networks do not experience frequent reconfiguration. For example, the closing of normally open switches happens off-load and very rarely. Furthermore, changes on cable sections, also rare, are planned in advance. Therefore, DNOs should have enough flexibility to update the polynomials when necessary. On the other hand, other expected network topologies could be captured in the offline simulations and implemented when necessary.

## 6. Conclusions

Lack of comprehensive and advanced telemetry in LV distribution feeders challenges the near-term implementation of centralized solutions for DG regulation. Combining offline simulations and regression analysis, the proposed method provides, to every DG unit, a series of polynomials that link feeder voltages, power/current flows and losses to local measurements without the need for remote monitoring. Essentially, the proposed polynomials can provide, in the near-term, observability to DG units at their location.

This paper focuses on describing the simulations required to obtain the mentioned polynomials and on the assessment of their adequacy for characterizing relevant network variables under multiple scenarios. It is shown how these polynomials can be implemented online to obtain accurate estimations, autonomously at each DG unit location, of the network variables based only on local measurements. Results show the methodology to be very accurate at estimating the magnitude of remote voltages, current/power flows and network losses. For instance, the estimation of voltages at the ending node of the feeder (a point that is likely to exceed upper voltage limits under a large presence of DG) presents an error of less than 2.5×10$^{-3}$ pu, this is regardless of the location of the DG unit which local measurements are being used for the estimation.

Differently from Distribution State Estimation, the proposed methodology provides with direct analytical expressions that characterise all network variables of interest as a function of the DG units power injections. This means that estimations are obtained through non-iterative calculations, presenting very low computational burden during real-time operation and no convergency risk. Furthermore, the polynomials do not only estimate the magnitude of the network variables but also describe the sensitivities of the latter to the controllable conditions at each DG unit location (i.e. the DG units active and reactive power injections). As the polynomials can estimate how much a network variable will change if the DG's power injections are changed (i.e. confining the network to a tractable constraint space that is a function of the controllable variables), they can directly enable fully decentralized control applications; an aspect that will be covered in future work.

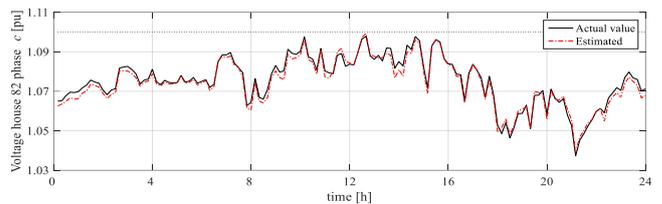

*Fig. 17.* House 82 phase c voltage – 50% PV and 30% EV penetration scenario

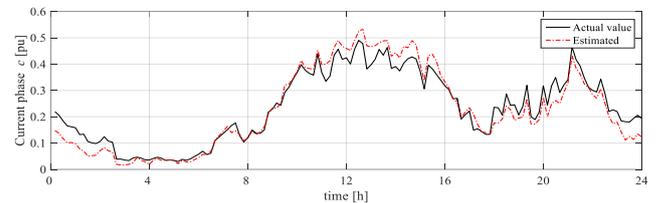

*Fig. 18.* Head of the feeder phase c current – 50% PV and 30% EV penetration scenario

## 7. Appendix

### 7.1. Benchmark Method: Decentralized Distribution State Estimation Algorithm

It is crucial to make a numerical comparison with other methods to benchmark the performance of the proposed methodology; for that reason, the estimations from the polynomials are benchmarked with a decentralized DSE algorithm. The proposed benchmark method is a LSE, i.e. the same formulation used in most State Estimation problems [11], [12]. In the LSE likelihood function (12), measurement errors are assumed to have a Gaussian distribution, with $E(z_i)$ and $\sigma_i$ being the expected value and the standard deviation of the ith measurement, $z_i$, respectively. As $E(z_i)$ is expressed as a non-linear function relating the system state variables to the i$^{th}$ measurement, LSE involves an iterative solution of a non-convex constraint optimization.

The proposed DSE (solved autonomously at each PV location) is formulated with the same monitoring restrictions as for the fitted polynomials, i.e. measurement of the local CPOC voltage and active and reactive power injections of the PV. Given that these measurements are insufficient for the condition of observability [11], the vector of measurements is increased with pseudo-measurements. Previous use of pseudo-measurements can be referred to the algorithm from [17], where they substitute a partial lack of real measurements. Here, a similar approach is followed where all remote demand measurements are replaced by statistical pseudo-measurements. In order to do that, the customers' expected active and reactive power demand and their standard deviations in (12) are derived from a pool of 10,000 demand profiles from the statistical model in [30]. In addition, local measurements are considered to have no error.

$$\min \sum_i^m \left(\frac{z_i - E(z_i)}{\sigma_i}\right)^2 \quad (12)$$

## 8. References

[1] R. A. Walling, R. Saint, R. C. Dugan, J. Burke, and L. A. Kojovic, "Summary of distributed resources impact on power delivery systems," *IEEE Trans. Power Deliv.*, vol. 23, no. 3, pp. 1636–1644, 2008.

[2] A. Alarcon-Rodriguez, E. Haesen, G. Ault, J. Driesen,

and R. Belmans, "Multi-objective planning framework for stochastic and controllable distributed energy resources," *IET Renew. Power Gener.*, vol. 3, no. 2, pp. 227–238, 2009.

[3] T. Stetz, K. Diwold, M. Kraiczy, D. Geibel, S. Schmidt, and M. Braun, "Techno-economic assessment of voltage control strategies in low voltage grids," *IEEE Trans. Smart Grid*, vol. 5, no. 4, pp. 2125–2132, 2014.

[4] Z. Zhang, L. F. Ochoa, and G. Valverde, "A Novel Voltage Sensitivity Approach for the Decentralized Control of DG Plants," *IEEE Trans. on Power Syst.*, vol. 33, no. 2, pp. 1566–1576, 2018.

[5] P. N. Vovos, A. E. Kiprakis, A. R. Wallace, and G. P. Harrison, "Centralized and distributed voltage control: Impact on distributed generation penetration," *IEEE Trans. Power Syst.*, vol. 22, no. 1, pp. 476–483, 2007.

[6] X. Su, M. A. S. Masoum, and P. Wolfs, "Comprehensive optimal photovoltaic inverter control strategy in unbalanced three-phase four-wire low voltage distribution networks," *IET Gener. Transm. Distrib.*, 2014.

[7] A. Rabiee, S. M. Mohseni-Bonab, M. Parniani, and I. Kamwa, "Optimal cost of voltage security control using voltage dependent load models in presence of demand response," *IEEE Trans. Smart Grid*, vol. 10, no. 3, pp. 2383–2395, 2019.

[8] A. R. Di Fazio, G. Fusco, and M. Russo, "Decentralized control of distributed generation for voltage profile optimization in smart feeders," *IEEE Trans. Smart Grid*, vol. 4, no. 3, pp. 1586–1596, 2013.

[9] Y. Wang, S. Wang, and L. Wu, "Distributed optimization approaches for emerging power systems operation: A review," *Electric Power Syst. Research*, vol. 144. pp. 127–135, 2017.

[10] K. Tanaka *et al.*, "Decentralised control of voltage in distribution systems by distributed generators," *IET Gener. Transm. Distrib.*, vol. 4, no. 11, pp. 1251–1260, 2010.

[11] A. Abur and A. Gomez Exposito, *Power System State Estimation: Theory and Implementation*. 2004.

[12] K. Dehghanpour, Z. Wang, J. Wang, Y. Yuan, and F. Bu, "A survey on state estimation techniques and challenges in smart distribution systems," *IEEE Trans. on Smart Grid*, vol. 10, no. 2, pp. 2312-2322, 2018.

[13] P. Siano, "Demand response and smart grids - A survey," *Renew. Sustain. Energy Rev.*, vol. 30, pp. 461–478, 2014.

[14] M. E. Baran and A. W. Kelley, "State Estimation for Real-Time Monitoring of Distribution Systems," *IEEE Trans. Power Syst.*, vol. 9, no. 3, pp. 1601–1609, 1994.

[15] M. E. Baran and A. W. Kelley, "A branch-current-based state estimation method for distribution systems," *IEEE Trans. Power Syst.*, vol. 10, no. 1, pp. 483–491, 1995.

[16] A. Angioni, T. Schlösser, F. Ponci, and A. Monti, "Impact of pseudo-measurements from new power profiles on state estimation in low-voltage grids," *IEEE Trans. Instrum. Meas.*, vol. 65, no. 1, pp. 70–77, 2016.

[17] R. Singh, B. C. Pal, and R. A. Jabr, "Distribution system state estimation through Gaussian mixture model of the load as pseudo-measurement," *IET Gener. Transm. Distrib.*, vol. 4, no. 1, pp. 50–59, 2010.

[18] D. A. Haughton and G. T. Heydt, "A Linear State Estimation Formulation for Smart Distribution Systems," *IEEE Trans. Power Syst.*, vol. 28, no. 2, pp. 1187–1195, 2013.

[19] M. C. De Almeida and L. F. Ochoa, "An Improved Three-Phase AMB Distribution System State Estimator," *IEEE Trans. Power Syst.*, vol. 32, no. 2, pp. 1463–1473, 2017.

[20] A. Alimardani, F. Therrien, D. Atanackovic, J. Jatskevich, and E. Vaahedi, "Distribution System State Estimation Based on Nonsynchronized Smart Meters," *IEEE Trans. on Smart Grid*, vol. 6, no. 6. pp. 2919–2928, 2015.

[21] A. T. Procopiou and L. F. Ochoa, "Voltage Control in PV-Rich LV Networks Without Remote Monitoring," *IEEE Trans on Power Systems*, vol. 32, no. 2. pp. 1224–1236, 2017.

[22] V. Rigoni and A. Keane, "Remote voltage estimation in LV feeders with local monitoring at transformer level," in *IEEE PES General Meeting*, 2018.

[23] C. Murphy and A. Keane, "Local and Remote Estimations using Fitted Polynomials in Distribution Systems," *IEEE Trans on Power Syst.*, vol.32, no.4. p.1, 2016.

[24] V. Rigoni and A. Keane, "Estimation of voltage sensitivities in low voltage feeders with photovoltaics," in *IEEE PES ISGT-Europe 2018*, 2018.

[25] G. W. Milligan and M. C. Cooper, "A study of standardization of variables in cluster analysis," *J. Classif.*, vol. 5, no. 2, pp. 181–204, 1988.

[26] A. K. Jain and R. C. Dubes, "Algorithms for Clustering Data," *Prentice Hall*, vol. 355. p. 320, 1988.

[27] S. C. Chapra and R. P. Canale, *Numerical methods for engineers*, vol. 33, no. 3. 2015.

[28] J. W. Smith, W. Sunderman, R. Dugan, and B. Seal, "Smart inverter volt/var control functions for high penetration of PV on distribution systems," in *IEEE PES PSCE*, 2011.

[29] Electric Power Research Institute - EPRI, "Common Functions for Smart Inverters: 4th Edition," 2016.

[30] K. McKenna and A. Keane, "Residential Load Modeling of Price-Based Demand Response for Network Impact Studies," *IEEE Trans. on Smart Grid*, vol. 7, no. 5. pp. 2285–2294, 2016.

[31] ENWL, "LV network models," *Low Voltage Network Solutions*, 2014.

[32] V. Rigoni, L. F. Ochoa, G. Chicco, A. Navarro-Espinosa, and T. Gozel, "Representative residential LV feeders: A case study for the North West of England," *IEEE Trans. Power Syst.*, vol. 31, no. 1, pp. 348–360, 2016.

[33] Energy Saving Trust, "Solar Energy Calculator Sizing Guide," 2011.

[34] ESB Networks Limited, "Distribution Code," 2016.

[35] EPRI, "OpenDSS - Open Distribution System Simulator," *smartgrid.epri.com*, 2014. .

[36] The Mathworks Inc., "MATLAB - MathWorks," *www.mathworks.com/products/matlab*, 2016.

[37] V. V Petrov, *Limit theorems of probability theory: Sequences of independent random variables*, vol. 4. 1995.

[38] G. Chicco, "Overview and performance assessment of the clustering methods for electrical load pattern grouping," *Energy*, vol. 42, no. 1, pp. 68–80, 2012.